\documentclass[aip,reprint]{revtex4-1}
\usepackage{amsmath}
\usepackage{verbatim}

\usepackage{float}
 \usepackage{graphicx}
 \usepackage{epstopdf}
 \usepackage{color}

\begin{document}

\newcommand{\Ql}{Q_{\mathrm{l}}}
\newcommand{\Qc}{Q_{\mathrm{c}}}
\newcommand{\Qi}{Q_{\mathrm{i}}}
\newcommand{\Qe}{Q_{\mathrm{e}}}
\newcommand{\fr}{f_\mathrm{r}}
\newcommand{\K}{\mathrm{K}}
\newcommand{\mK}{\mathrm{mK}}
\newcommand{\Tc}{T_{\mathrm{c}}}
\newcommand{\Hz}{\mathrm{Hz}}
\newcommand{\kHz}{\mathrm{kHz}}
\newcommand{\MHz}{\mathrm{MHz}}
\newcommand{\GHz}{\mathrm{GHz}}
\newcommand{\St}{S_{21}}
\newcommand{\Tone}{T_1}
\newcommand{\s}{\mathrm{s}}
\newcommand{\ms}{\mathrm{ms}}
\newcommand{\us}{\mu\mathrm{s}}
\newcommand{\ns}{\mathrm{ns}}
\newcommand{\mus}{\mu\mathrm{ s}}
\newcommand{\eV}{\mathrm{eV}}
\newcommand{\V}{\mathrm{V}}
\newcommand{\mV}{\mathrm{mV}}
\newcommand{\uV}{\mu\mathrm{V}}
\newcommand{\nA}{\mathrm{nA}}
\newcommand{\m}{\mathrm{m}}
\newcommand{\mm}{\mathrm{mm}}
\newcommand{\um}{\mu \mathrm{m}}
\newcommand{\nm}{\mathrm{nm}}
\newcommand{\cm}{\mathrm{cm}}
\newcommand{\Watt}{\mathrm{W}}
\newcommand{\dBm}{\mathrm{dBm}}
\newcommand{\Torr}{\mathrm{Torr}}
\newcommand{\mTorr}{\mathrm{mTorr}}
\newcommand{\dB}{\mathrm{dB}}

\newcommand{\Ohm}{\Omega}
\newcommand{\kohm}{\mathrm{k}\Omega}

\newcommand{\F}{\mathrm{F}}
\newcommand{\pF}{\mathrm{pF}}
\newcommand{\fF}{\mathrm{fF}}
\newcommand{\Henry}{\mathrm{H}}
\newcommand{\nH}{\mathrm{nH}}
\newcommand{\Tesla}{\mathrm{T}}
\newcommand{\mT}{\mathrm{mT}}

\newcommand{\M}{\mathrm{M}}
\newcommand{\nph}{\left\langle n_\mathrm{ph}\right\rangle}
\newcommand{\kB}{k_\mathrm{B}}

\newcommand{\Pin}{P_{\mathrm{in}}}
\newcommand{\abs}[1]{\left| #1 \right|}

\title{Reducing intrinsic loss in superconducting resonators by surface treatment and deep etching of silicon substrates}

\author{A. Bruno}
\affiliation{QuTech Advanced Research Center and Kavli Institute of Nanosicence, Delft University of Technology, Lorentzweg 1, 2628 CJ Delft, The Netherlands}

\author{G. de Lange}
\affiliation{QuTech Advanced Research Center and Kavli Institute of Nanosicence, Delft University of Technology, Lorentzweg 1, 2628 CJ Delft, The Netherlands}

\author{S. Asaad}
\affiliation{QuTech Advanced Research Center and Kavli Institute of Nanosicence, Delft University of Technology, Lorentzweg 1, 2628 CJ Delft, The Netherlands}

\author{K. L. van der Enden}
\affiliation{QuTech Advanced Research Center and Kavli Institute of Nanosicence, Delft University of Technology, Lorentzweg 1, 2628 CJ Delft, The Netherlands}

\author{N. K. Langford}
\affiliation{QuTech Advanced Research Center and Kavli Institute of Nanosicence, Delft University of Technology, Lorentzweg 1, 2628 CJ Delft, The Netherlands}

\author{L. DiCarlo}
\affiliation{QuTech Advanced Research Center and Kavli Institute of Nanosicence, Delft University of Technology, Lorentzweg 1, 2628 CJ Delft, The Netherlands}

\date{\today}

\begin{abstract}
We present microwave-frequency NbTiN resonators on silicon, systematically achieving internal quality factors above 1~M in the quantum regime.
We use two techniques to reduce losses associated with two-level systems: an additional substrate surface treatment prior to NbTiN deposition to optimize the metal-substrate interface, and deep reactive-ion etching of the substrate to displace the substrate-vacuum interfaces away from high electric fields.
The temperature and power dependence of resonator behavior indicate that two-level systems still contribute significantly to energy dissipation, suggesting that more interface optimization could further improve performance.
\end{abstract}

\maketitle

Superconducting coplanar waveguide (CPW) microwave resonators are crucial elements in photon detectors~\cite{Day03}, quantum-limited parametric amplifiers~\cite{Castellanos-Beltran08, Bergeal10} and narrow-band filters~\cite{Endo13}, as well as read-out, interconnect and memory elements in quantum processors based on circuit quantum electrodynamics~\cite{Blais04}.
They also play a critical role in hybrid devices, connecting superconducting circuits with micro- and nanomechanical resonators~\cite{Regal08, Teufel08} and solid-state spins~\cite{Kubo10,Amsuss11}.
In many quantum science and technology applications, resonators must operate in the quantum regime, requiring low temperatures to reach the ground state (thermal energy $\kB T$ small compared to the photon energy at resonance, $h f_r$) and single-photon excitation levels.
Under these conditions, however, internal quality factors ($\Qi$s) are typically substantially lower than their high-temperature or high-power values.

In the quantum regime, the dominant loss mechanism for high-$Q$ superconducting resonators can be attributed to parasitic two-level systems (TLSs) in the dielectrics~\cite{Gao08, OConnell08}.
TLSs may reside in the bulk substrate~\cite{OConnell08}, as well as in the metal-substrate, metal-vacuum and substrate-vacuum interfaces~\cite{Gao08, Vissers10, Barends08, Barends10a, Megrant12, Weber11, Khalil11, Geerlings12c} where electric fields may be large (see Ref.~\onlinecite{Oliver13} and references therein for a recent review of material-related loss in superconducting circuits).
Interface TLSs are common by-products of the fabrication process, often introduced by impurities associated with substrate surfaces~\cite{Wisbey10, Quintana14} and etching chemistry~\cite{Chen08}. 
To our knowledge, the best resonators reported to date~\cite{Megrant12} ($\Qi=1.72~\mathrm{M}$ at 6~GHz) are fabricated by epitaxially growing aluminum on sapphire substrates following careful surface preparation (high-temperature annealing in an oxygen atmosphere).
For CPW resonators on silicon (Si) substrates, achieving $\Qi>1~\M$ in the quantum regime has proven challenging, with the best resonators reported in Ref.~\onlinecite{Sandberg12}.

\begin{figure}[t]	
	\includegraphics[width=\columnwidth]{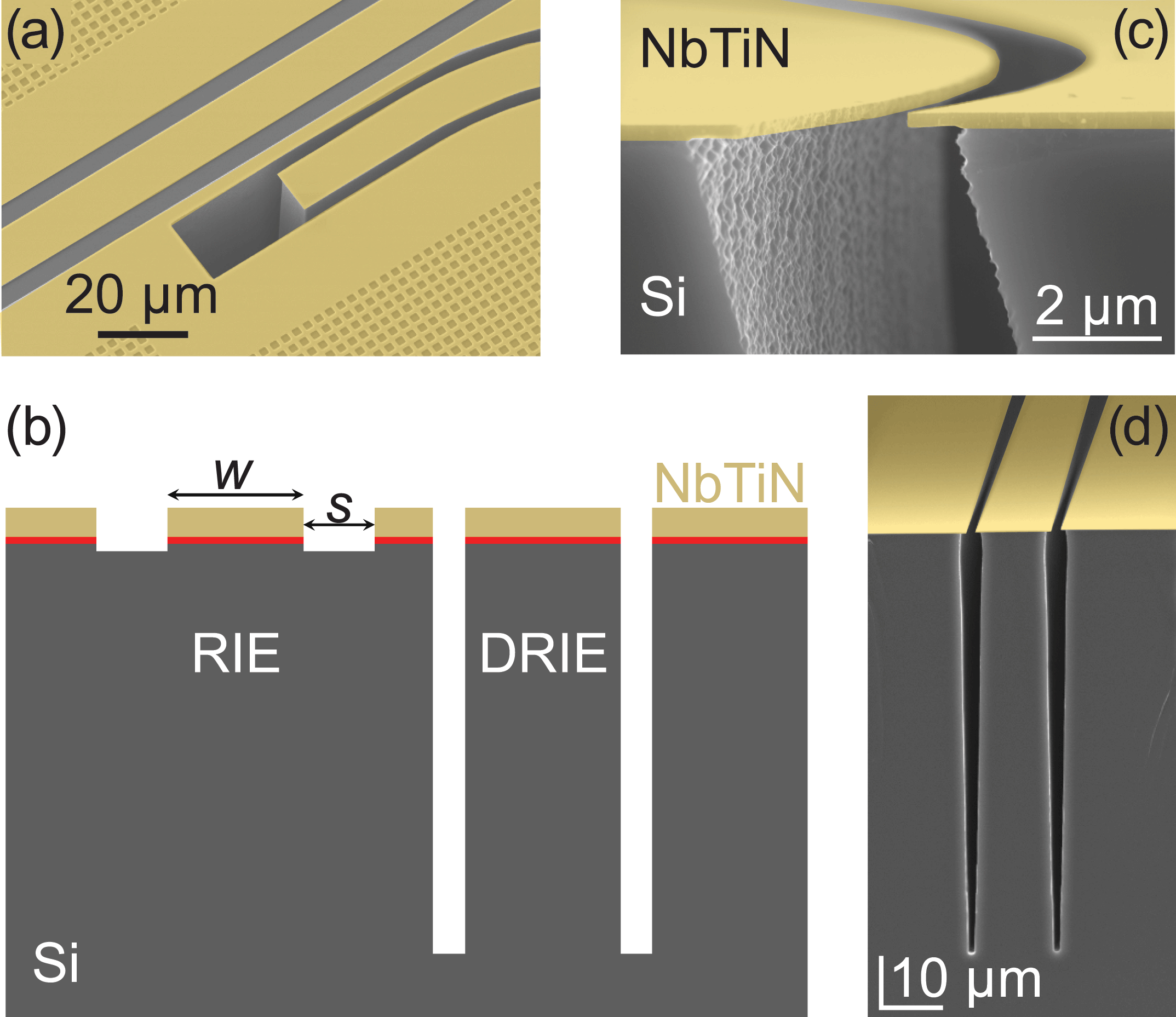}
		\caption{False-colored scanning electron micrographs of CPW resonators and their cross sections. (a) Capacitive coupling of a RIE feedline to a DRIE resonator. (b) Schematic cross section of RIE and DRIE CPW resonators (vertical axis not to scale) in a NbTiN thin film (gold shading) on a Si substrate (grey). The treated substrate-metal interface is highlighted in red. RIE (DRIE) etches $\sim150~\nm$  $(\sim80~\um)$ into the substrate. (c) Cross section of a DRIE resonator near the surface, showing the metallization ($300~\nm$), Si underetch ($\sim 1~\um$) and sidewall roughness ($\sim 100~\nm$).
(d) Cross section of DRIE resonator, showing the high etch anisotropy and depth ($\sim 80~\um$).}
\label{fig:Schematics}
\end{figure}

In this letter, we present silicon-based, gigahertz-frequency CPW resonators with $\Qi$ systematically above $1~\M$ in the quantum regime, fabricated from niobium titanium nitride (NbTiN) superconducting films.
This performance is reached by optimizing two aspects of the fabrication.
First, the substrate surface is treated with hexamethyldisilazane (HMDS) immediately prior to metal deposition to reduce losses associated with the metal-substrate interface.
Second, we employ highly anisotropic deep reactive-ion etching (DRIE) of the substrate to displace lossy substrate-vacuum interfaces away from regions of high electric field.
We obtain the best performance by combining both techniques, reaching $2~\M$ in the quantum regime.

Resonator fabrication starts with surface preparation of high-resistivity ($\rho = 150~\kohm\cm$) Si $\langle$100$\rangle$ substrates for deposition of the superconducting film.
We choose NbTiN as the superconductor for its high-quality metal-vacuum interface~\cite{Barends10a}. 
Because we define the resonators via etching of the metal film rather than lift-off, we can apply surface treatments immediately prior to metal deposition to optimize the substrate-metal interface.
The substrate surface quality depends on its physical condition (e.g., roughness), chemical composition (e.g., presence of native oxides or nitrides), and chemical cleanliness (e.g., absence of polymers or metals).
For example, thin ($\sim 2~\nm$) amorphous native silicon oxide layers are known to host a considerable density of TLSs~\cite{OConnell08,Barends08}.
Various adsorbates can in turn attach to the native oxides.
For instance, hydroxyl groups (-OH) couple to dangling bonds on the substrate surface and build a layer of silanol groups (Si-OH). This silanol layer is also hydrophilic, and thus adsorbs additional water and dipolar molecules.  
We first remove lossy oxides from the surface following the standard practice of dipping the Si substrate in a 1:7 buffered solution of hydrofluoric acid (HF) for $2~\min$, which both etches away surface oxides and terminates the Si surface with hydrogen.
We then include an additional step immediately following the HF dip, placing the substrate on a hot plate for $2~\min$ at $110^{\circ}\mathrm{C}$ while exposing the surface to an HMDS-nitrogen(N$_2$) atmosphere. 
HMDS is an organosilicon compound most commonly used as a primer to improve resist adhesion, to create hydrophobic surfaces, and to prepare substrates in the growth of high-mobility graphene~\cite{Lafkioti10}.
XPS surface analysis of substrates immediately after these treatments revealed an absence of Si-O bonds for both the HF-only and HF+HMDS cases~\cite{SOMapl}.
After also analyzing the dynamics of surface re-oxidation~\cite{SOMapl}, we conclude that the Si surface does not re-oxidize either during the HMDS treatment or loading of the substrate in the high-vacuum chamber for NbTiN deposition.
Following surface preparation, NbTiN is sputtered in a reactive atmosphere of argon (Ar) and N$_2$ at $3~\mTorr$ (5~\% atm -of N$_2$) and $400~\Watt$ (DC). The films are 160 or $300~\nm$ thick~\cite{SOMapl}, with a typical critical temperature of $15.5~\K$ and $\rho=110~\mu\Omega\cm$.

We next aim to reduce intrinsic losses associated with the substrate-vacuum interface.
We employ deep reactive-ion etching (DRIE) of the exposed Si substrate using the Bosch process~\cite{Bosch96} to move the substrate-vacuum interface away from regions of high electric field (Fig.~1a-b).
The Bosch process, consisting of alternating etching and passivation steps, is highly tunable, with variable etch depth, under-etch, sidewall roughness and passivation~\cite{Zuwei13}.
The DRIE resonators presented here have $\sim1~\um$ under-etch below the edges of the metal regions, side-wall scalloping with $\sim100~\nm$ features (Fig.~1c), and a PTFE fluorocarbon layer uniformly deposited on the exposed Si surfaces during the etching process.

We now describe a recipe which allows fabrication of both RIE and DRIE structures on the same chip.
First, the NbTiN films are patterned using e-beam lithography and RIE to define feedlines and standard resonators.
Next, an 8-$\um$-thick layer of AZ~9216 photoresist is spun, followed by $25~\nm$ of sputtered tungsten and a layer of PMMA~A4.
After e-beam patterning and developing the PMMA, a short RIE step is used to etch the tungsten, creating a metallic mask for a $1~\mathrm{h}$ anisotropic $\mathrm{O}_2$-plasma ashing of the AZ resist.
We then deep-etch the Si substrate, alternating between etching using an inductively-coupled plasma (ICP) of SF$_6$ gas ($20~\mTorr$, $2200~\Watt$, $7~\s$) and passivation with an ICP of C$_4$F$_8$ ($2~\mTorr$, $1200~\Watt$, $2~\s$).
The thick AZ layer allows etching to a depth of $80~\um$ (Fig.~1d) and $500~\nm$ minimum feature size.

For the devices in this study, we employ quarter-wave CPW resonators with center conductor width fixed to $W = 12~\um$ and separation $S$ between the center conductor and ground planes set to give a transmission-line impedance close to $50~\Omega$.
For RIE resonators, we set $S=5~\um$, taking into account $\epsilon_r = 11.9$ for the silicon substrate and a kinetic inductance fraction of $\sim 16\%$.
For DRIE resonators, we decreased $S$ to $2~\um$ to compensate for the reduced effective dielectric constant.
From resonator frequencies, we estimate a DRIE resonator impedance of $\sim62~\Omega$.
The resonators are capacitively coupled to a common feedline with a target coupling quality factor $\Qc$ between $700~\K$ and $1~\M$. 
After dicing and final cleaning in NMP, the devices are mounted in a copper PCB with Apiezon N grease, wire-bonded and anchored to the mixing chamber plate of a dilution refrigerator routinely employed for characterizing resonators (and qubits) in the quantum regime~\cite{SOMapl}.

\begin{figure}[t]
	\includegraphics[width=\columnwidth]{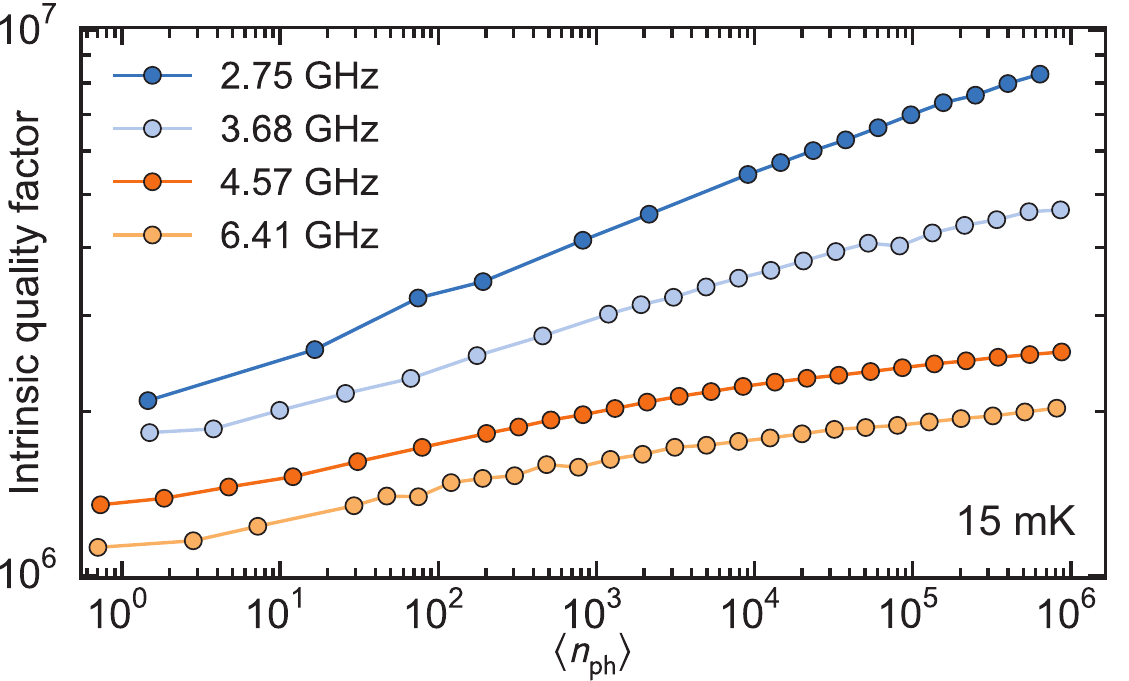}
	\caption[Needs editing]{Intrinsic quality factors of four DRIE resonators measured at 15 mK. $\Qi$ is extracted from measurements of complex-valued feedline transmission.  The DRIE resonators (NbTiN thickness $\sim 300~\um$, central conductor width $W=12~\um$, and gap $S=2~\um$) have fundamental frequencies ranging from 2.75 to 6.41~GHz.  The best-fit values of $\fr$, $\Qi$ and $\Qc$ are used to convert the calibrated input power to $\nph$~[\onlinecite{SOMapl}]. All resonators show $\Qi > 1~\M$ in the quantum regime, with higher $\Qi$s attainable for higher $\nph$ and lower $\fr$.  Error bars are smaller than the symbol size.}
	\label{fig:Fig2}
\end{figure}

We quantify resonator performance by measuring their power and temperature-dependent resonance lineshapes.
Specifically, we measure the complex-valued feedline transmission $S_{21}$ near the fundamental frequencies $\fr$ of each coupled resonator, and fit the data~\cite{Khalil12,SOMapl} to extract $\fr$ and $\Qi$.
Here, we report measurements for our best resonators, fabricated using both surface treatment and DRIE techniques (see supplementary material~\cite{SOMapl} for other devices).
We show $\Qi$ as a function of excitation power for four HMDS-DRIE resonators with fundamental frequencies $\fr = 2.75-6.41~\GHz$ (Fig.~2).
All resonators show $\Qi>1~\M$ in the quantum regime (average intraresonator photon number $\nph \sim 1$, $T \sim 15~\mK$).
However, we observe further increases in $\Qi$ (up to 4 times at 2.75 GHz) as the resonator is populated with more photons~\cite{Barends10}.
This suggests that the low-power limits for the resonators are still significantly influenced by TLSs.
We also observe a systematic increase in $\Qi$ with decreasing $\fr$, explanations for which include the decrease in TLS polarization at lower frequencies and the decrease in radiation losses at longer wavelengths. 

\begin{figure}[t]
	\includegraphics[width=\columnwidth]{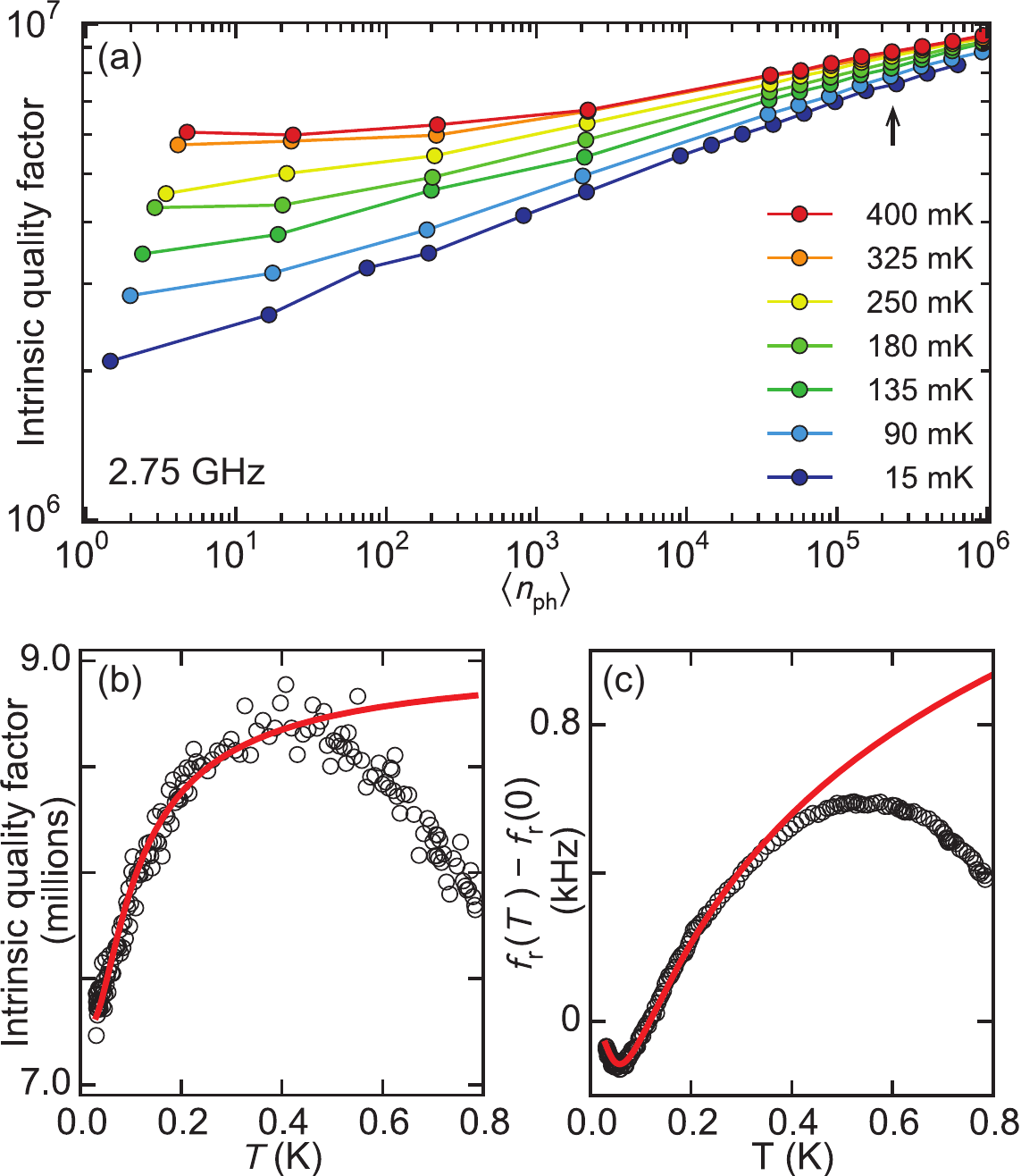}
	\caption[Needs editing]{Temperature dependence of $2.75~\GHz$ DRIE resonator.  (a) Power dependence of $\Qi$ measured for $T = 15$--$400~\mK$. At high powers, $\Qi$ is similar for all $T$, but temperature dependence increases for lower $\nph$. (b,c) Resonator probed at fixed $\nph \sim 10^5$ (black arrow in (a)) while device is cooled from $900$ to $15~\mK$ (plotted against mixing chamber temperature).  Both $\Qi$ (b) and $\Delta f$ (c) show non-monotonic temperature dependence.  We attribute the initial increase in $\Qi$ and $\Delta f$ (as temperature decreases) to a decrease in quasiparticle-related loss. At $T\lesssim400~\mK$, $\Qi$ and $\Delta f$ begin to decrease as TLS-related losses increase.  Data fitting to a simple model~\cite{Pappas11} for $\Delta f$ (solid red line in (c); Eq.~1) shows that, in this regime, the behaviour is dominated by the $T$-dependence of TLS polarization.  Modifying the related TLS model for $\Qi$ to account for high excitation power (solid red line in (b); Eq.~2) provides estimates for a TLS polarization fraction at $\nph \sim 10^5$ of $\sim 4~\%$, and a background (non-TLS) loss of $\sim1/(9~\M)$ which is consistent with high-power $\Qi$s from (a).}
	\label{fig:Fig3}
\end{figure}

To study the effects of TLS polarization, we acquire the power dependent $\Qi$ of the $2.75~\GHz$ resonator for $T = 15$--$400~\mK$ (Fig.~3a).
For all excitation powers, $\Qi$ increases monotonically with temperature over this range, and higher temperatures make $\Qi$ less power dependent, with the factor of 4 change at $15~\mK$ dropping to a factor of only 1.5 at around $400~\mK$. 
We therefore conclude that TLSs still provide significant loss in the quantum regime, and attribute the increase of $\Qi$ with temperature to thermal depolarization of TLSs.

A more detailed temperature dependence is acquired by measuring $S_{21}$ around $2.75~\GHz$ at fixed $\nph \sim 10^5$ while the device is continuously cooled from $900$ to $15~\mK$.
Fitting $S_{21}$ as before, we extract $\Qi$ and $\fr$.
Initially, $\Qi$ increases during cooling due to decreasing quasiparticle-induced loss~\cite{Barends08}, before reaching a maximum value of $\sim 8.6~\M$ at $T\sim0.4~\K$ (Fig.~3b).
Below $0.4~\K$, $\Qi$ decreases again as increasing TLS polarization leads to greater absorption of resonator photons. 

Resonator loss from TLSs predominantly depends on their density of states at $\fr$ and the applied microwave power~\cite{Oliver13,Phillips87}.
The continuing power dependence of $\Qi$ at single-photon levels (Fig.~2) shows that even at these low powers, resonant TLSs are still being depolarized by the incident microwaves, suggesting a significant loss contribution from TLSs with low saturation powers or, equivalently, long lifetimes.
Because $\Qi$ does not reach its low-power baseline, this limits the usefulness of the $\Qi$ measurements in quantifying the TLS-related loss. 
An alternative way to probe the effect of TLSs is to study the temperature dependence of $\fr$.
Unlike loss, the dispersive shift of the resonator also depends on non-resonant TLSs that are not saturated by the applied microwaves~\cite{Pappas11}.
Their influence on $\fr$ can therefore be measured even at high $\nph$.
Figure~3c shows the temperature dependence of the resonator frequency shift $\Delta \fr(T) = \fr(T) - \fr(0)$ extracted from the same data as Fig.~3b.
The local maximum in $\Delta \fr$ at $T \sim 500~\mK$ again reflects the loss-dependent pull of the resonator, but unlike $\Qi$, $\Delta \fr$ is not monotonic below this temperature.

The low-temperature behavior of $\Delta \fr$ is well described by a two-parameter TLS model~\cite{Pappas11}
\begin{align}
 \nonumber
&\Delta\fr(T) = \\
 &\frac{k}{\pi}\left[ \mathrm{Re}~\Psi\left(\frac{1}{2}{+}\frac{1}{2\pi i}\frac{h\fr(T)}{\kB T}\right) \right.
-\left. \ln\left(\frac{1}{2\pi}\frac{h\fr(T)}{\kB T}\right)\right]\fr(0).
\end{align}
Here, $\Psi$ is the complex digamma function, and $\fr(0)$ is the resonator frequency at $T=0$, and $k = F\delta$ is a combined loss parameter, where $\delta$ is the loss tangent (at $T=0$ and $\nph = 0$) and $F$ the filling factor of the TLS host material.
Fitting the TLS model with $k$ and $\fr(0)$ as free parameters (red curve in Fig.~3c) shows good agreement with the data for $T<400~\mK$ and accurately captures the distinctive local minimum in $\Delta f$ at $ T\sim 60~\mK$.
The best-fit value $k = 6.35(4)\cdot 10^{-7}$ compares well with previous values found for Nb resonators on sapphire~\cite{Gao08}.
This value can also be used to predict the low-power, low-temperature limit of TLS-related loss\cite{Pappas11}, giving $\Qi \approx 1/k \sim 1.6~\M$.
The $\sim20~\%$ disparity with the measured single-photon $\Qi$ (Fig.~2) fits with previous observations~\cite{Pappas11}.  It arises partly because $\Delta \fr$ is sensitive to TLSs over a broad frequency bandwidth, giving an estimate of average loss tangent $\delta$ which may differ from the local loss tangent inside the resonator bandwidth~\cite{Pappas11}, and partly because we do not reach a baseline value of $\Qi$ even at the single-photon level.

We also fit an heuristic model for the low-temperature dependence of $\Qi$ modified from Ref.~\onlinecite{Pappas11}:
\begin{align}
\frac{1}{\Qi} = k_{\rm eff} \tanh\left( \frac{h\fr(T)}{2 \kB T} \right) + \frac{1}{Q_{\mathrm{other}}}.
\end{align}
Here, we introduce two \emph{ad hoc} fitting parameters, a modified loss parameter $k_{\rm eff}$ to account for the reduced TLS polarization at high excitation power, and an extra, fixed loss term $1/Q_{\rm other}$ which accounts for the non-TLS-related loss which becomes significant as temperature increases.
This fit also agrees well with the data at $T \lesssim 0.4~\K$ (red curve in Fig.~3b), with best-fit values of $k_{\rm eff} = 2.64(4) \cdot 10^{-8}$ and $Q_{\rm other} = 9.02(2)\cdot10^6$.
The ratio $k_{\rm eff}/k_{\rm fit} \sim 0.04$ provides an estimate of the TLS polarization fraction still remaining at $\nph \sim 10^5$, while $Q_{\rm other}$ is consistent with the high-power $\Qi$s measured in Fig.~3a.

In conclusion, we have demonstrated few-GHz CPW resonators on silicon with intrinsic quality factors exceeding one million in the quantum regime, achieved through surface treatment and substrate etching.
Our results suggest that these resonators exhibit a significant reduction in TLS-related losses, but the temperature and power dependences indicate that TLSs are still an important source of energy loss.
This offers the opportunity for further improvements by interface optimization, perhaps pushing quantum-regime resonator performance towards the high-power limit with $\Qi$s approaching 10~M at lower frequencies.
Since we have only partially explored the available parameter ranges in this study, future efforts will include optimizing the DRIE recipe and studying the effects of etch depth, side-wall roughness and under-etch on resonator losses.
Finally, these surface-treatment and DRIE techniques can be directly applied to the fabrication of superconducting qubits in integrated circuits, offering a new path towards longer qubit coherence times.\\

\begin{acknowledgments}

We thank T.~M.~Klapwijk, P.~J.~de~Visser and D.~J.~Thoen for discussions and high-quality NbTiN films, and
V.~de Rooij  and E. van Veldhoven for the XPS investigation.
We acknowledge funding from the Dutch Organization for Fundamental Research on Matter (FOM)
and a Marie Curie Career Integration Grant (L.D.C.).

\end{acknowledgments}


\clearpage
\onecolumngrid

\begin{appendix}
\renewcommand{\theequation}{S\arabic{equation}}
\renewcommand{\thefigure}{S\arabic{figure}}
\setcounter{figure}{0}

\section*{Supplemental Material}

This supplement provides experimental details and data sets supporting claims made in the main text.
First, we present design and fabrication details for three device generations that we have measured: Device C (reported in the main text) and devices from Sets A and B (reported below).
We then describe several technical details of the measurement system: the measurement setup and device shielding, the fitting of the complex-valued transmission spectra, and the procedure for estimating $\nph$.
Next, we present measurement results for devices from Sets A and B.
Finally, we discuss the results of XPS analysis of the pre-deposition substrate surface following different treatments.

\section*{Devices}

\begin{figure*}[!htb]
	\includegraphics[width=480 pt]{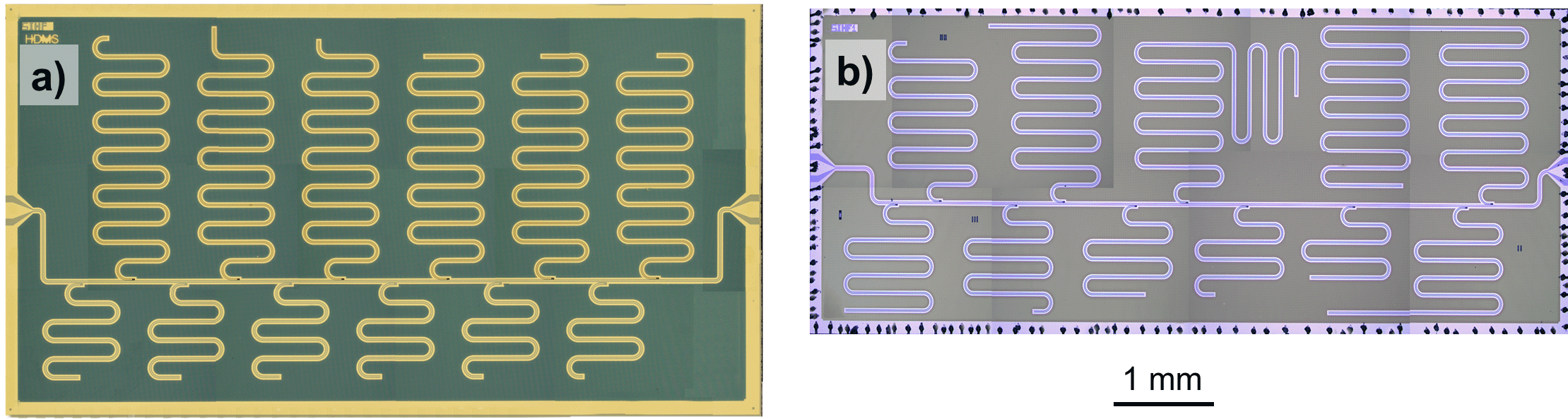}
	\caption[Needs editing]{a) Optical image of HMDS-treated device from Set~B (Set B used the same layout as Set A). The feedline and the lower 6 resonators are etched by RIE while the top 6 resonators are patterned by DRIE.  Frequencies for the RIE and DRIE resonators are centered around $7$ and $4~\GHz$, respectively. b) Optical image of Device C (HMDS+DRIE).  The feedline is patterned using RIE, but all resonators are fabricated using DRIE with frequencies spanning 2 to $11~\GHz$.}
	\label{fig:DRIE_optical}
\end{figure*}

Devices in Sets A and B had identical layouts~(Fig.~\ref{fig:DRIE_optical}a).
Each device contained 6 RIE resonators with frequencies $\sim7~\GHz$ (spaced by $\sim 100~\MHz$; see results in Figs.~\ref{fig:supp-results-A-RIE} and \ref{fig:supp-results-B-RIE}) and 6 DRIE resonators with frequencies $\sim4~\GHz$ (also spaced by $\sim 100~\MHz$; see results in Figs.~\ref{fig:supp-results-A-DRIE} and \ref{fig:supp-results-B-DRIE}).
In each set, we fabricated one device using the standard HF-only substrate treatment (green dashed curves in Figs.~\ref{fig:supp-results-A-RIE}--\ref{fig:supp-results-B-DRIE}) and one device using the HF+HMDS substrate treatment (solid red curves in Figs.~\ref{fig:supp-results-A-RIE}--\ref{fig:supp-results-B-DRIE}).
In Sets A and B, the DRIE recipe is slightly different from the recipe described in the main text (for Device C).
For these sets, the plasma power used in the DRIE etching step was set at 1800~W.
At the end of the deep etch process, the samples underwent an additional etch with an SF$_6$ inductively coupled plasma (20 mTorr, 1800 W, 60 s with no alternating passivation step).
This removed the PTFE passivation and created a slightly isotropic etch profile.
The fabrication was otherwise almost identical for Sets A and B, with the NbTiN film $\sim 160~\nm$ thick and the DRIE etch $\sim 80~\um$ deep.
The only difference was that the devices in Set~B were fabricated with the inclusion of a short (20 sec) additional fluorocarbon (PTFE) side-wall passivation step in the DRIE recipe after the isotropic etch, while no such final step was included in the devices from Set A.

Device C (recipe and data in main text) only included DRIE resonators fabricated using the combined HF+HMDS surface treatment.
We used a thicker ($\sim 300~\nm$) NbTiN film, and kept the DRIE etch depth the same ($\sim 80~\um$).
The device had 10 resonators with frequencies between $2$ and $11~\GHz$.
The plasma power in the etching step in the DRIE recipe was increased to 2200~W to reduce the feature size of the side-wall scalloping.

\newpage
\section*{Measurement Setup}

\begin{figure}[!htb]
	\includegraphics[width=230 pt]{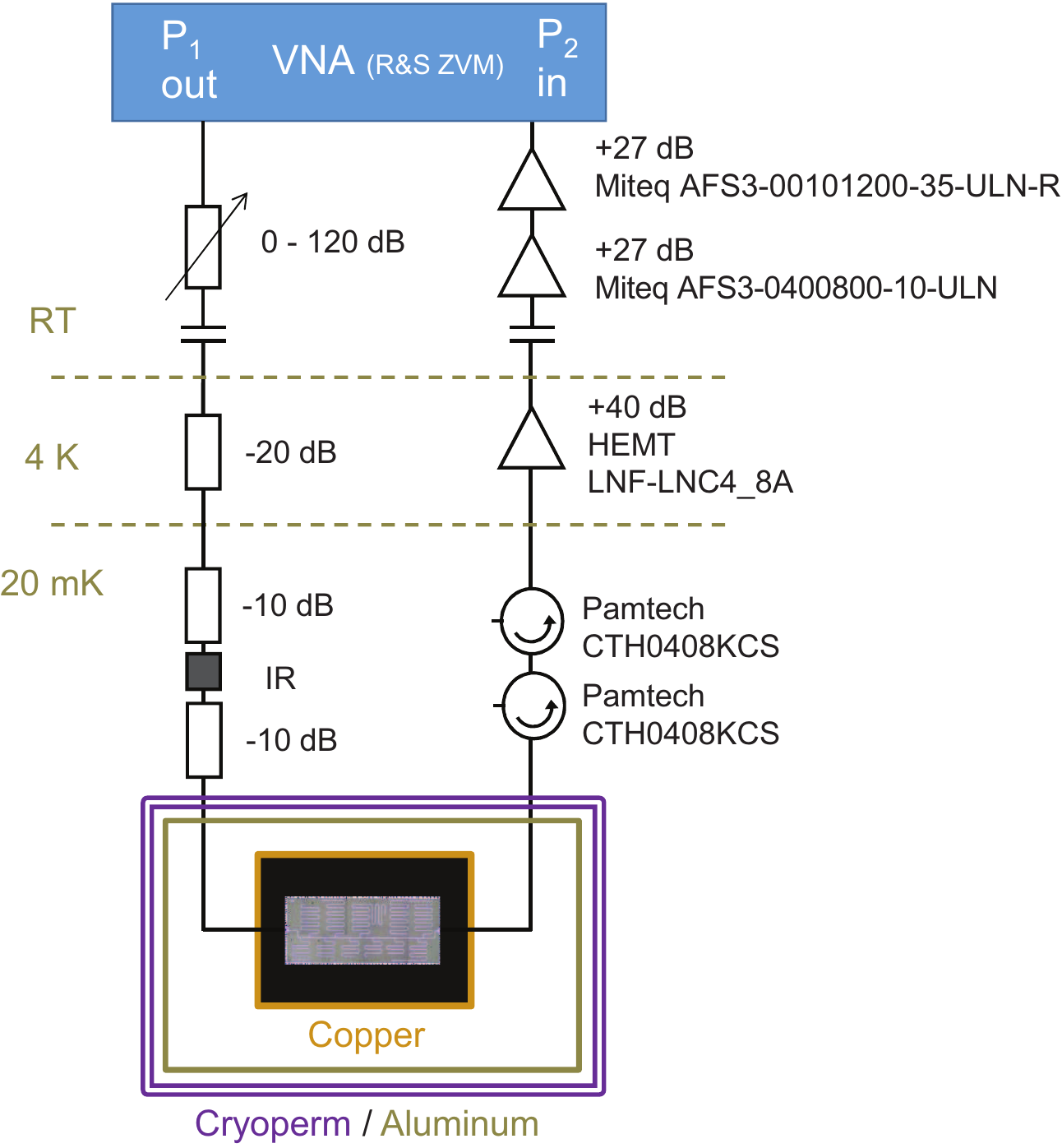}
	\caption[width=\textwidth]{Measurement setup used for resonator characterization in the $^{3}\mathrm{He}/^{4}\mathrm{He}$ dilution refrigerator (Leiden Cryogenics CS81): microwave schematic, magnetic shields (two Cryophy, one aluminum), IR filters (home-made Eccosorb filters), and IR adsorbers (inner surface of the inner Cu radiation shield, coated with Stycast~2850 and silicon carbide powder).}
	\label{fig:fridges}
\end{figure}

In an effort to minimize resonator losses induced by non-equilibrium quasiparticles~\cite{Barends11} and magnetic vortex displacement~\cite{Bothner11}, we mount the samples inside several layers of shielding~(Fig.~\ref{fig:fridges}).
Specifically, we anchor the PCB-mounted sample directly to a copper cold finger connected to the mixing chamber of the dilution refrigerator.
The sample is then enclosed in a copper can, the inner surface of which is coated in a mixture of Stycast~2850 and silicon carbide granules with diameters between $15$ and $1000~\um$.
This can is enclosed in a second can made from aluminum.
This is finally enclosed in two layers of cryogenic magnetic shielding (1-mm-thick Cryophy, Magnetic Shields Ltd).
Coaxial cables enter the sample chamber through tubulations in the lid of the outer layer of magnetic shielding to reduce the impact of open holes on the shielding effectiveness.
Extra radiation shielding is provided by home-made in-line Eccosorb filters in the input coaxial line mounted outside the magnetic shields at the mixing chamber stage.

\subsection*{Data fitting}

\begin{figure}[!htb]
	\includegraphics[width=0.8\columnwidth]{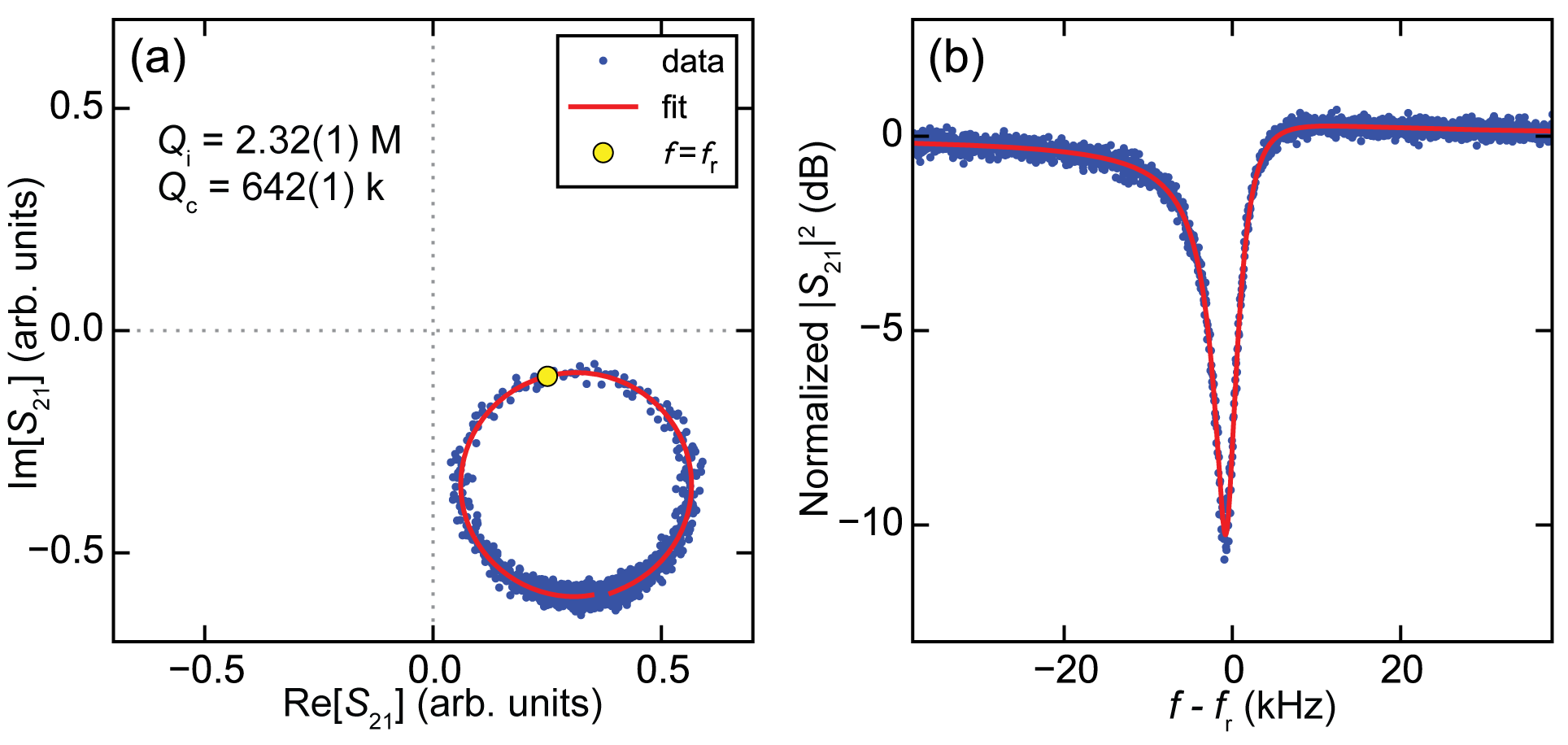}
	\caption[Needs editing]{Typical measured response of a CPW resonator around its resonance frequency, and fit. Data taken from DRIE resonator with $\fr = 3.68~\GHz$ at  $\nph \sim 10$~photons (Fig.~3 of main text). Panel (a) shows the trajectory of $\St$ in the complex plane with a fit (red line) of Eq.~\ref{eq:THEhanger}. (b) Normalized $\abs{S_{21}}$ calculated from the data and fit in (a). }
	\label{Fig:fitting}
\end{figure}

The transmission through the full measurement chain near one resonance is shown in Fig.~\ref{Fig:fitting}.
We fit the $S_{21}$ curves following the method presented in Ref.~\onlinecite{Khalil12}.  Equation~11 in Ref.~\onlinecite{Khalil12} describes the general form of a microwave resonator transmission measurement with a number of non-ideal elements, including both inductive and capacitive coupling, and impedance mismatch in the feedline:
\begin{equation}
\St =A\left(1-\frac{\frac{\Ql}{\lvert \Qe \rvert}e^{i\theta}}{1+2i\Ql\frac{f-\fr}{\fr}}\right).
\label{eq:THEhanger-original}
\end{equation}
Here, $\fr$ is the resonance frequency, $A$ is the transmission amplitude away from resonance, $\Ql$ is the loaded quality factor of the resonator, related to $\Qi$ and $\Qc$ by $1/\Ql = 1/\Qc + 1/\Qi$, and $\Qe = \left|\Qe\right| \exp(-i\theta)$ is a complex-valued quality factor related to $\Qc$ via $1/\Qc = \mathrm{Re}(1/\Qe)$.  The imaginary part of $1/\Qe$ gives rise to an asymmetry in the resonator lineshape.  We further modify Eq.~\ref{eq:THEhanger-original} as follows:
\begin{equation}
\St =A \left( 1 + \alpha \frac{f-\fr}{\fr} \right)
\left(1-\frac{\frac{\Ql}{\lvert \Qe \rvert}e^{i\theta}}{1+2i\Ql\frac{f-\fr}{\fr}}\right)e^{i\left(\phi_v f+\phi_0\right)}.
\label{eq:THEhanger}
\end{equation}
Here, $\alpha$ allows a linear variation in the overall transmission chain in the narrow frequency range around any given resonance, and $\phi_v$ and $\phi_0$ account for the propagation delay to and from the sample.
An example best fit of Eq.~\ref{eq:THEhanger} to data is also shown in Fig.~\ref{Fig:fitting}.

\subsection*{Calibration of the average number of photons in the resonator}

To estimate $\nph$, we calibrated the signal input line and the home-made eccosorb filter at low temperature in order to estimate the power $\Pin$ at the feedline input.
To calculate $\nph$ from $\Pin$, we note that the reflected and transmitted powers (ignoring feedline mismatch) are respectively given by:
\begin{align}
\nonumber
P_{\rm refl} &= \Pin |S_{11}|^2, \\
\nonumber
P_{\rm trans} &= \Pin |S_{21}|^2,
\end{align}
where $S_{21} = \Qc/(\Qc+\Qi)$ and $S_{11} = S_{21}-1 = -\Qi/(\Qc+\Qi)$ are the usual scattering parameters for a resonator-shunted feedline probed near resonance~\cite{barendsphd}.
The power absorbed in the resonator is therefore:
\begin{align}
\nonumber
P_{\rm abs} = \Pin -P_{\rm refl}-P_{\rm trans} = \frac{2\Ql^2}{\Qc\Qi}.
\end{align}
$P_{\rm abs}$ can also be written using the definition of internal loss rate $\kappa_{\rm i}$:
\begin{align}
\nonumber
P_{\rm abs} &= \nph \, \hbar \wr \, \kappa_{\rm i}, \\
\nonumber
&= \nph \, \frac{\hbar \wr^2}{\Qi}.
\end{align}
Combining these equations then gives the relation between $\nph$ and $\Pin$:
\begin{align}
\nonumber
\nph &= \frac{2}{\hbar\wr^2}\frac{\Ql^2}{\Qc}\Pin.
\label{eq:nphoton}
\end{align}
Due to the difficulty in accounting for all reflections in the full measurement line, these values can be interpreted as an estimated upper limit to $\nph$.

\clearpage
\section*{Results from Sets~A and B at 20 \MakeLowercase{m}K}

 \begin{minipage}{\linewidth}
   \centering
   \begin{minipage}{0.45\linewidth}
     \begin{figure}[H]
       \includegraphics[width=\linewidth]{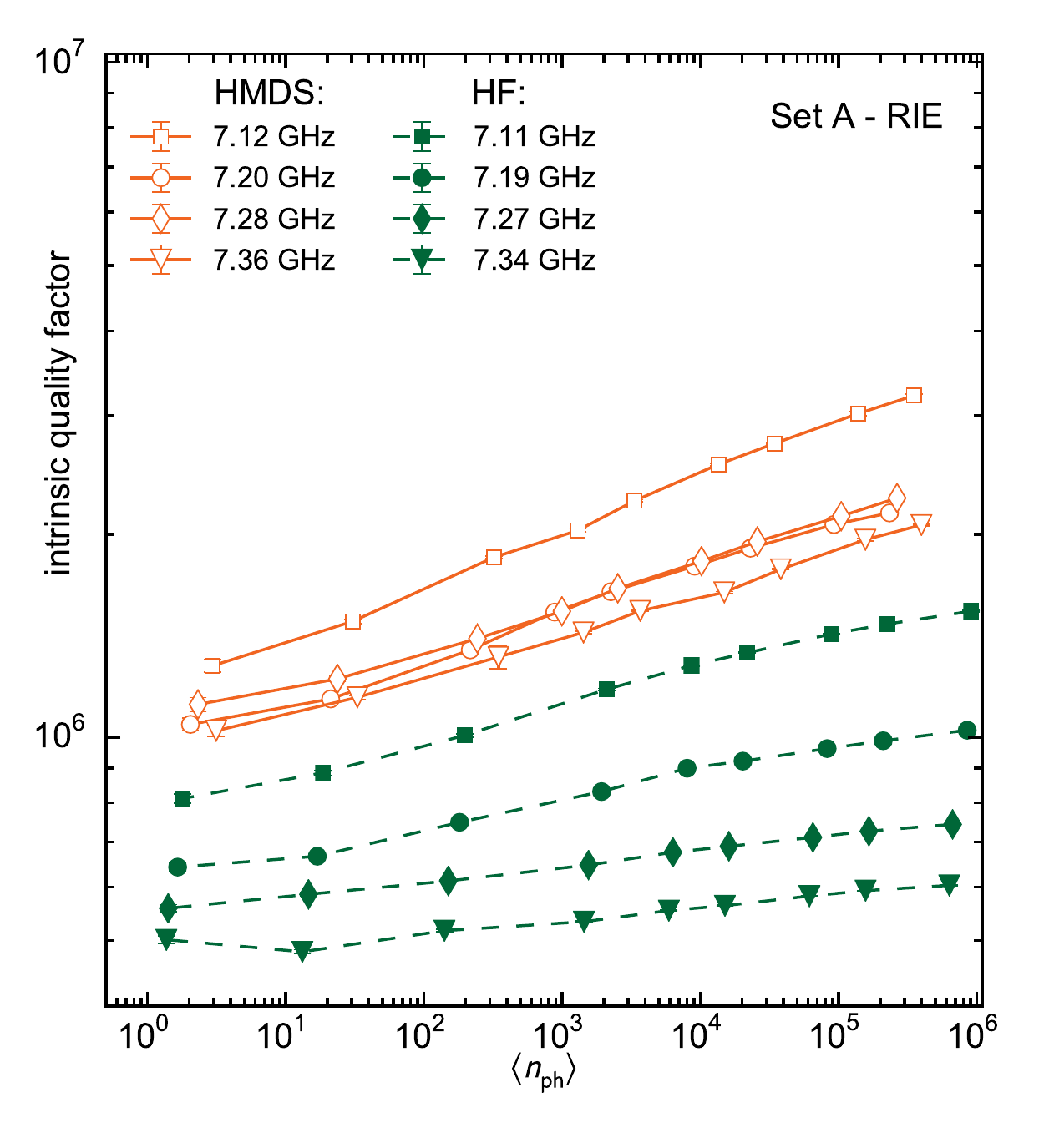}
     \caption[]{Comparison between HF and HMDS surface treatment of RIE resonators from Set~A.}
     \label{fig:supp-results-A-RIE}
       \end{figure}
 \end{minipage}
  \hspace{0.05\linewidth}
\begin{minipage}{0.45\linewidth}
   \begin{figure}[H]
      \includegraphics[width=\linewidth]{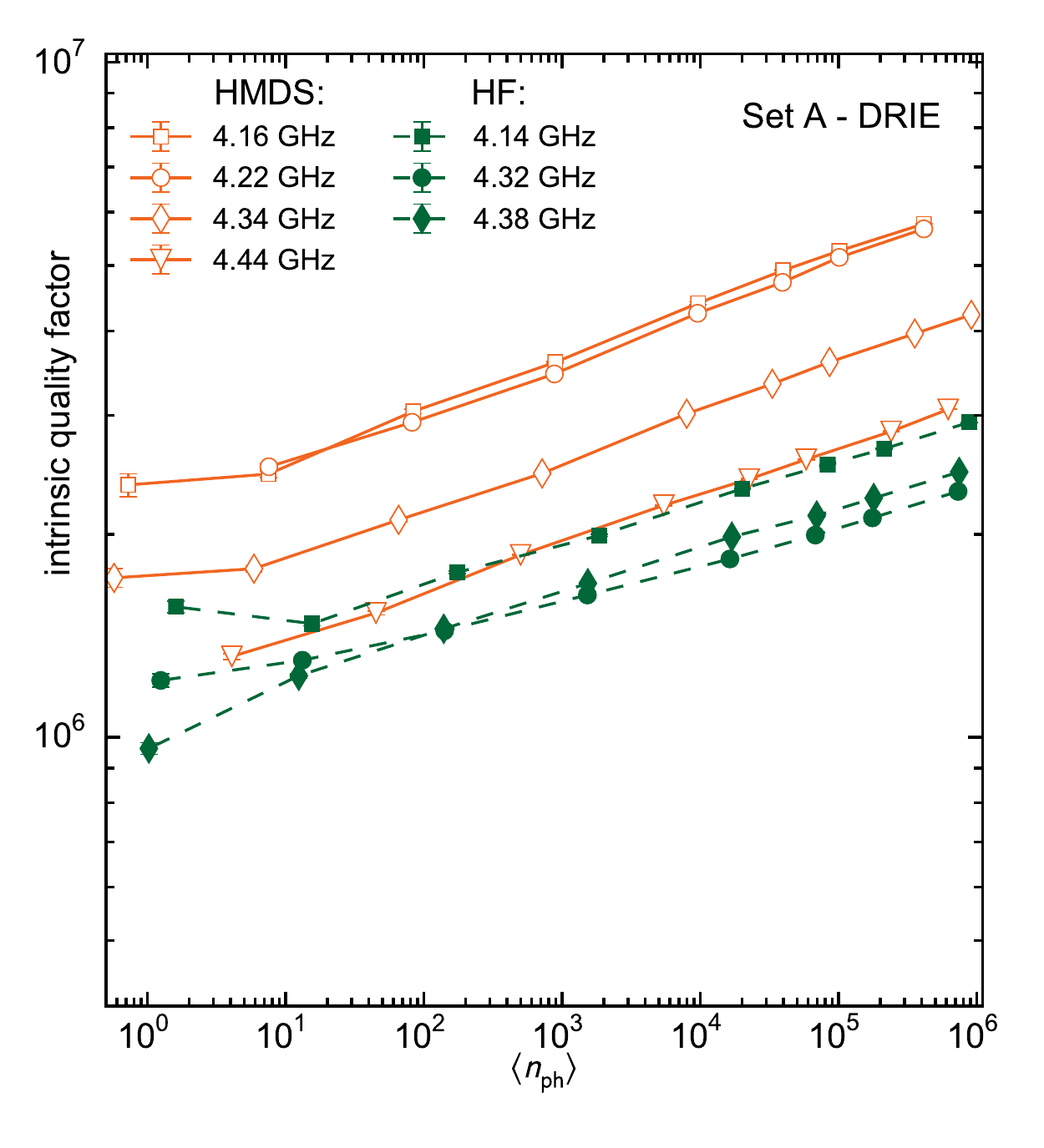}
    \caption[]{Comparison between HF and HMDS surface treatment of DRIE resonators from Set~A.
     \label{fig:supp-results-A-DRIE}
  }
\end{figure}
\end{minipage}
\end{minipage}

  \begin{minipage}{\linewidth}
      \centering
      \begin{minipage}{0.45\linewidth}
          \begin{figure}[H]
              \includegraphics[width=\linewidth]{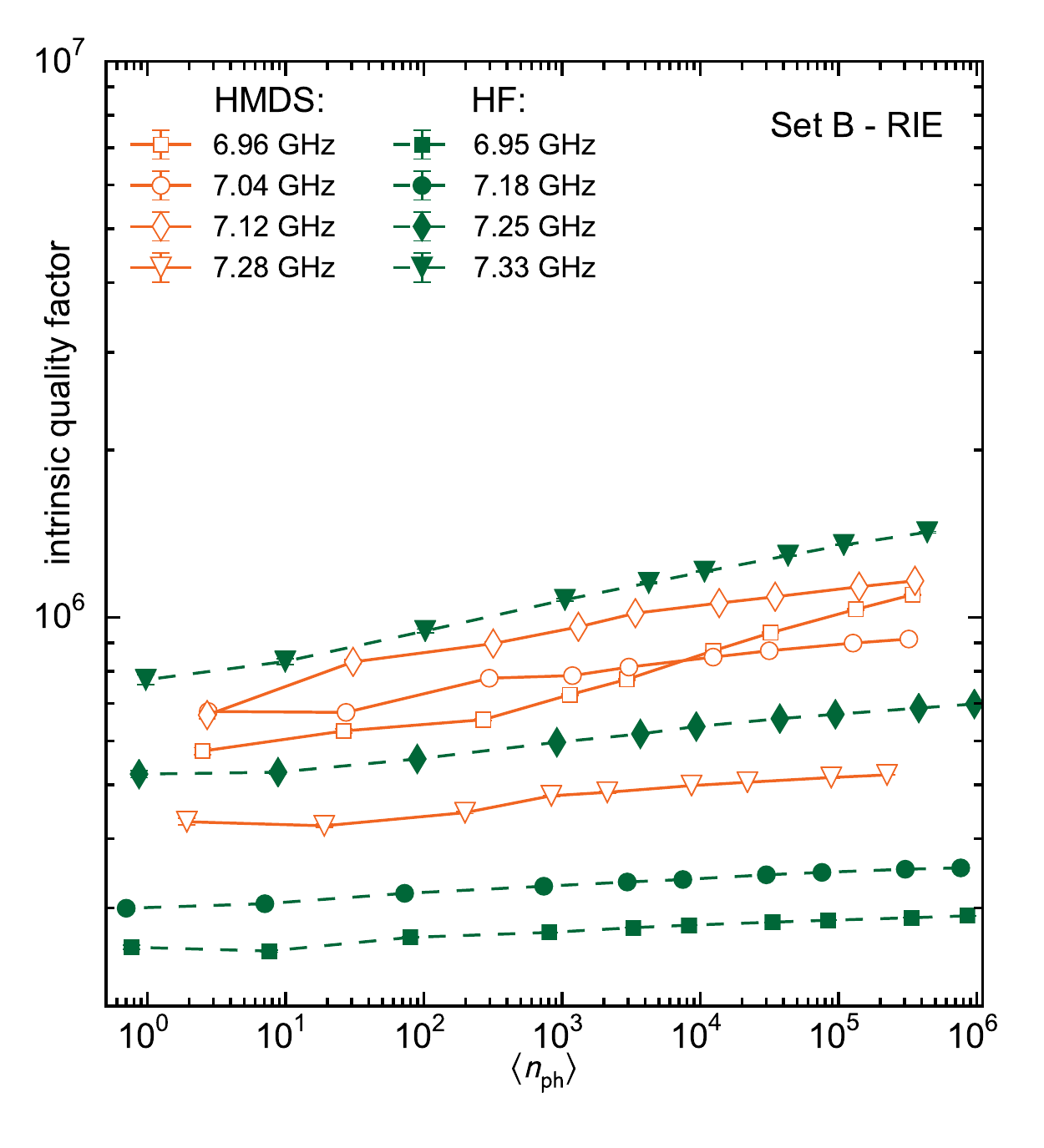}
              \caption[]{Comparison between HF and HMDS surface treatment of RIE resonators from Set~B.}
     \label{fig:supp-results-B-RIE}
          \end{figure}
      \end{minipage}
     \hspace{0.05\linewidth}
      \begin{minipage}{0.45\linewidth}
          \begin{figure}[H]
              \includegraphics[width=\linewidth]{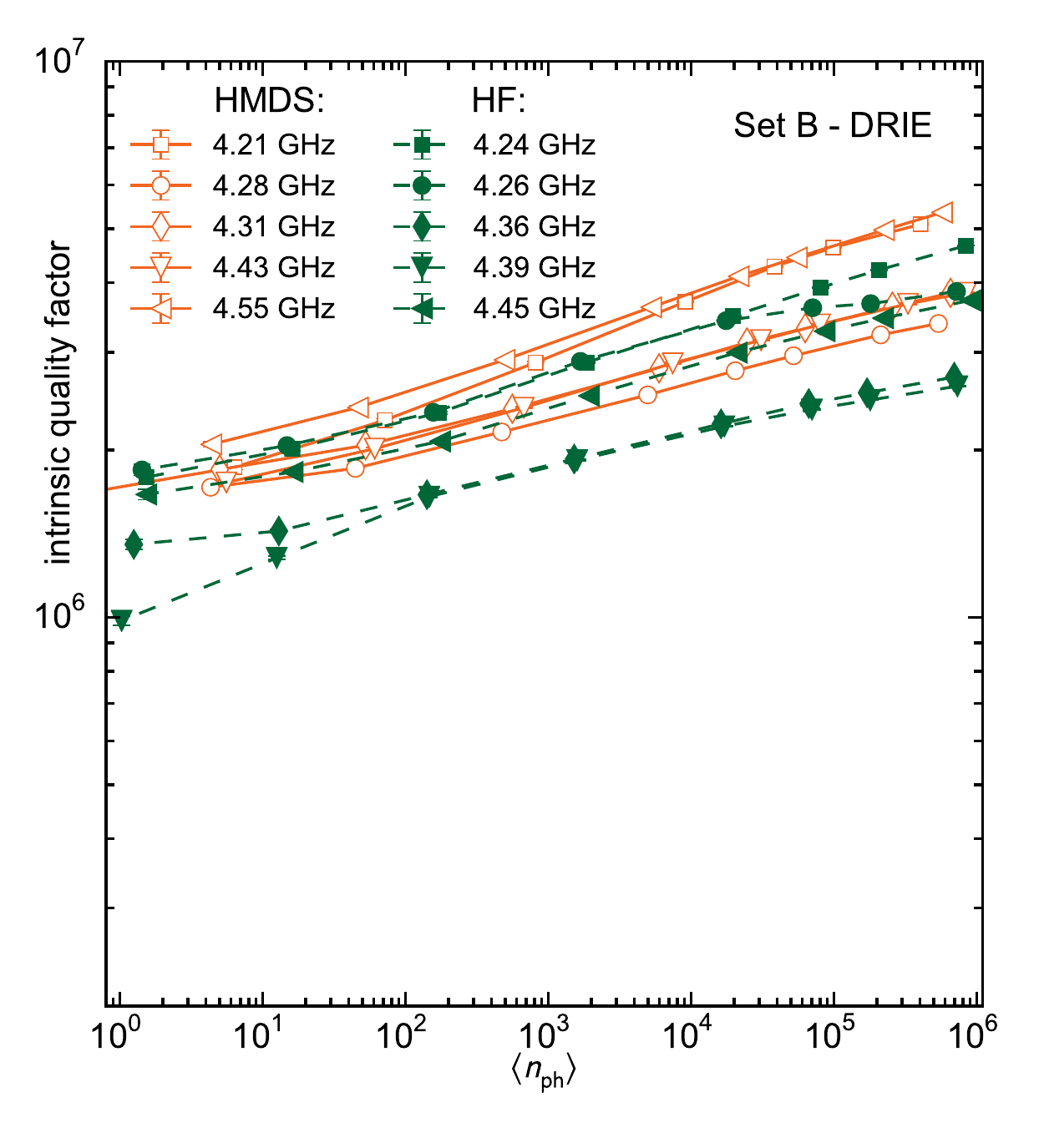}
              \caption[]{Comparison between HF and HMDS surface treatment of DRIE resonators from Set~B.}
     \label{fig:supp-results-B-DRIE}
          \end{figure}
      \end{minipage}
  \end{minipage}

\section*{XPS data}

We performed XPS surface analysis of Si substrates in order to estimate surface oxygen uptake during the HMDS treatment following the HF dip, as well as during the approximately two-minute interval between finishing the surface preparation and loading the sample into the high-vacuum chamber ($P_{\mathrm{base}}= 2\times10^{-8}~\mathrm{Torr}$) for NbTiN deposition.

\begin{figure}[ht!]
              \includegraphics[width=\linewidth]{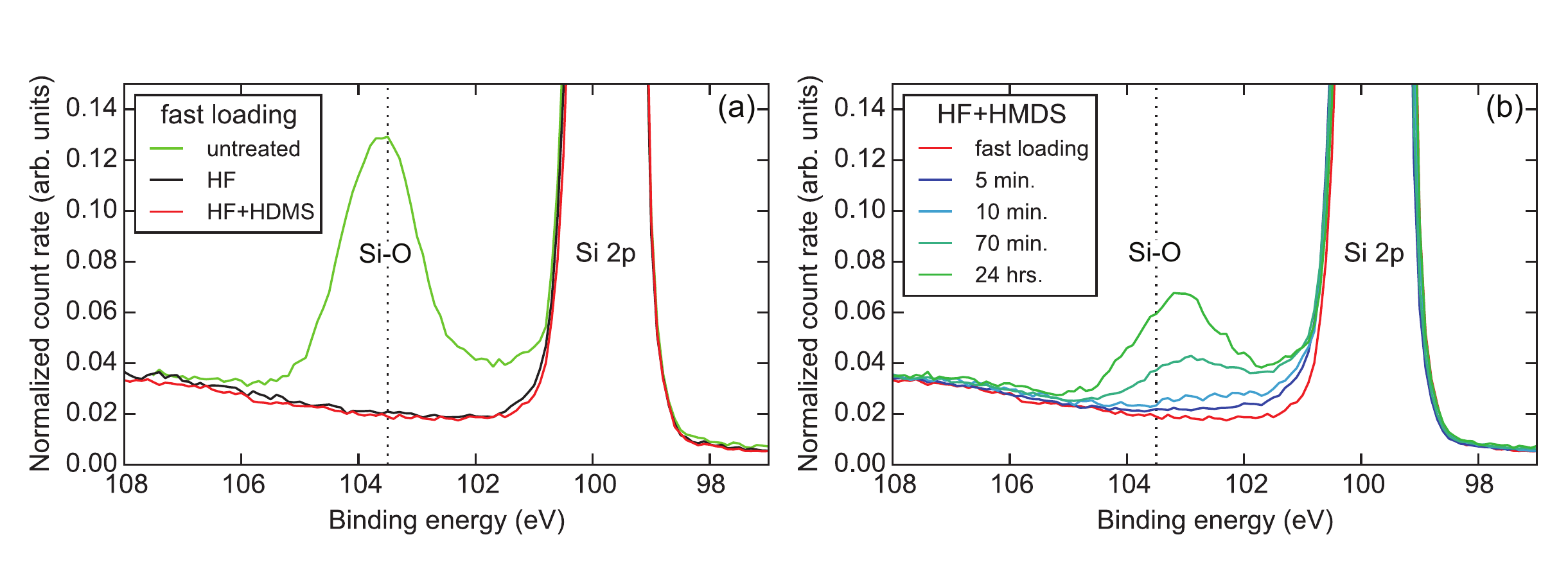}
              \caption[]{XPS analysis after fast loading of substrates ($\sim 2~\min$ after surface processing). (a) XPS spectrum of the Si-O peak at $103.5\pm0.3~\eV$ for: an untreated Si surface with native oxides (green), a Si surface treated by HF dip only (black), and a Si surface treated by HF dip and HMDS bubbling (red) as described in the main text. (b) Time evolution of the XPS Si-O peak for a Si surface treated by HF dip and HMDS bubbling after subsequent exposures to atmospheric conditions (times indicated in the legend). (We thank V.~de Rooij and E.~van Veldhoven for carrying out the XPS measurements.)}
	\label{fig:xps}
 \end{figure}

Figure~\ref{fig:xps}a shows that the standard HF surface treatment already removes all signatures of SiO from the substrate surface.  No discernible change is observed when substrates are placed in a mixed nitrogen-HMDS atmosphere, created by HMDS bubbling, following the HF dip.
Figure~\ref{fig:xps}b shows the XPS spectrum after subsequent exposures to air.  While measurement directly after fast loading does not reveal a Si-O peak, subsequent exposures of the sample to atmospheric conditions show the onset of one, indicating oxygen adsorption by the substrate surface.

\section*{Supplementary References}

\end{appendix}

\end{document}